\documentclass[conference]{IEEEtran}
\IEEEoverridecommandlockouts
\usepackage{subfigure}
\usepackage{multirow}
\usepackage{booktabs}
\usepackage{cite}
\usepackage{amsmath,amssymb,amsfonts}
\usepackage{algorithmic}
\usepackage{graphicx}
\usepackage{textcomp}
\usepackage{xcolor}

\usepackage{wrapfig}
\usepackage{float}
\usepackage{comment}

\usepackage[hyphens]{url}
\usepackage[hidelinks]{hyperref}
\hypersetup{breaklinks=true}
\usepackage{cite}

\def\BibTeX{{\rm B\kern-.05em{\sc i\kern-.025em b}\kern-.08em
    T\kern-.1667em\lower.7ex\hbox{E}\kern-.125emX}}
\begin{document}

\title{An artificial neural network-based system for detecting machine failures using tiny sound data: A case study \\

}

\makeatletter
\newcommand{\linebreakand}{%
  \end{@IEEEauthorhalign}
  \hfill\mbox{}\par
  \mbox{}\hfill\begin{@IEEEauthorhalign}
}
\makeatother

\author{\IEEEauthorblockN{Anonymous Authors}}
\author{
    \IEEEauthorblockN{Thanh Tran\IEEEauthorrefmark{1}, Sebastian Bader\IEEEauthorrefmark{1}, Jan Lundgren\IEEEauthorrefmark{1}}
    \IEEEauthorblockA{\IEEEauthorrefmark{1} STC Research Centre, Department of Electronics Design, Mid Sweden University
    \\\{thanh.tran, sebastian.bader, jan.lundgren\}@miun.se, thanhttp02@gmail.com}}

\maketitle

\begin{abstract}
In an effort to advocate the research for a deep learning-based machine failure detection system, we present a case study of our proposed system based on a tiny sound dataset. Our case study investigates a variational autoencoder (VAE) for augmenting a small drill sound dataset from Valmet AB. A Valmet dataset contains 134 sounds that have been divided into two categories: "Anomaly" and "Normal" recorded from a drilling machine in Valmet AB, a company in Sundsvall, Sweden that supplies equipment and processes for the production of biofuels. Using deep learning models to detect failure drills on such a small sound dataset is typically unsuccessful. We employed a VAE to increase the number of sounds in the tiny dataset by synthesizing new sounds from original sounds. The augmented dataset was created by combining these synthesized sounds with the original sounds. We used a high-pass filter with a passband frequency of 1000 Hz and a low-pass filter with a passband frequency of 22\kern 0.16667em000 Hz to pre-process sounds in the augmented dataset before transforming them to Mel spectrograms. The pre-trained 2D-CNN Alexnet was then trained using these Mel spectrograms. When compared to using the original tiny sound dataset to train pre-trained Alexnet, using the augmented sound dataset enhanced the CNN model's classification results by 6.62\%(94.12\% when trained on the augmented dataset versus 87.5\% when trained on the original dataset). 
\end{abstract}


\section{Introduction}
\IEEEPARstart{T}{he} drilling machines are commonly used in factories to drill holes in materials. In this paper, a failure detection system for Valmet AB's drilling machines is investigated using Valmet AB's drilling sounds. A drilling machine operated by Valmet AB includes many drill bits used for drilling thousands of small holes in metal surfaces. Drill bits that break can cause serious damage to a product since they lead to a lack of holes in the metal surface. As a result, a technician manually punches holes into metal sheets that are missing holes. It is a very time-consuming and expensive process. Hence, technicians usually check the drilling machine every 10 minutes so that if any drill bit breaks they can replace it before the drilling machine continues. Furthermore, shutting the machine off and on every 10 minutes is labor-intensive and time-consuming. Hence, the factory would benefit from an automated system that can classify drilling sounds and notify technicians when drill bits break. Our machine failure detection proposal builds on the fact that skilled technicians can detect a drill that isn't working properly by listening to its sound when cutting metal.

Typically, large, balanced datasets were used to build a machine failure detection system. Data anomalies related to the damaged machine in Valmet AB are probably only a small proportion of the total data set. As a result of the imbalance of real-world datasets, CNN tends to be biased; for instance, CNN will struggle to predict minority classes if they have fewer data. The collection and labeling of sounds can be a time-consuming and expensive process for a company. These operational costs can be reduced by using data augmentation techniques to transform datasets. Additionally, in order to create variations that are representative of what a deep learning model might see in reality, data augmentation techniques enable deep learning models to be more robust.

Sound applications use standard augmentation methods for extending small datasets. There are classic and advanced techniques in sound augmentation. The conventional approach of sound augmentation methods includes time stretching, pitch shifting, volume control, adding noise, time-shifting, etc \cite{noauthor_augment_nodate}. Research in \cite{tran2021detecting} showed that some augmentation methods are not appropriate to apply in our short sounds (20.83 ms or 41.67 ms). The advanced approach for data augmentation is using Synthetic Minority Over-sampling Technique (SMOTE) \cite{chawla_smote:_2002} or  Modified Synthetic Minority Over-sampling Technique (MSMOTE)\cite{5403368}, etc. Overall, data augmentation helps to increase the number of sounds in the dataset. This improves the prediction accuracy of deep learning models and reduces data overfitting. Additionally, it increases the generalization of deep learning models by creating variable data. 

Recently, deep learning augmentation methods such as generative adversarial network (GAN) ~\cite{8698632,9006268,9411996,9390834,8615782,8985892,8717183, 8962219, 8857905} have been widely used as data augmentation methods in computer vision. However, GANs require large amounts of training data to generate effective augmented data.
Variational Autoencoder (VAE) was proposed in 2014 by Kingma and Welling \cite{kingma2014autoencoding}. VAE has been used in domain adaption in face recognition frameworks~\cite{9432318}, network representation learning~\cite{9449483}, detect anomaly in log file systems~\cite{8944863}, statistically extract latent space hidden in sampled data to control the robot~\cite{9143393}, to generate more data for software fault prediction~\cite{8672311}, traffic anomaly detection in videos~\cite{9531567}, for time series anomaly detection~\cite{9053558}, etc.
VAE and different improved versions of VAE have been used to generate synthetic data in remote sensing images~\cite{9393473}, for phrase-level melody representation learning to generate music~\cite{9054554}, to generate sensor data for chiller fault diagnosis~\cite{9264218}, to generate conditional handwritten characters~\cite{9511404, 8803129}, etc.
The purpose of this study is to generate new drilling sounds using VAE. To the best of our knowledge, this is the first time VAE was investigated from a drilling sound augmentation standpoint for the machine failure detection system. 

The classification of sound signals generated by machines has been investigated in recent years. In conventional machine learning techniques, statistical features from acoustic signals are extracted and classified~\cite{Lee2016},~\cite{Kemalkar2017},~\cite{Zhang2019a}. In machine failure analysis, machine learning algorithms are commonly used because of their robustness and adaptability on small datasets~\cite{9252126}. These conventional machines learning methods  include Support Vector Machine (SVM)~\cite{fengqi_compound_2006,Li2011TheAO,deng_novel_2019}, k‑Nearest Neighbors~\cite{7360161, WANG2016201, GLOWACZ201965}, decision tree~\cite{saravanan_fault_2009}, ANN and radial basis function
(RBF)~\cite{1265797,10.1016/j.neucom.2011.03.043, doi:10.1177/1077546313518816}, etc.

The use of deep learning to detect mechanical faults by analyzing acoustic signals has been successfully applied~\cite{polat2020fault,zhang2020deep,islam2018motor,verstraete2017deep,chen2017vibration,Long2020a,Paul2019,Luo2018,Ince2016}. According to recent studies, deep learning architectures can be trained to identify sound signals by using image representation, such as Mel frequency cepstral coefficients (MFCCs)~\cite{Zhang2017,8635051}, spectrogram~\cite{Boddapati2017}, Mel spectrogram~\cite{Mushtaq2021,9252126}, log-Mel spectrogram~\cite{tran2021detecting}. 

When detecting failures on rotating machines, a conventional approach of handcrafted feature extraction and selection is often used. However, extracting and selecting the right features can be challenging in order to use in a classifier or to accurately detect faults in an industrial setting. It is possible for both sound and noise to occur simultaneously in a realistic soundscape. Additionally, the drill sound waveform is complex and short, making it difficult to detect (around 20.83~ms and 41.67~ms). It is therefore not guaranteed that the extracted features of raw audio signals are sufficient for classification. 

The use of vibration sensors is common when detecting machine faults. On the contrary, our methods of detecting broken drills were based on sound signals for these reasons. Drill bits for each drilling machine in Valmet AB are available in quantities of 90 or 120. A vibration sensor is required for each drill bit in order to detect fractures. Mounting 90 to 120 vibration sensors on a drilling machine simultaneously and classifying these vibrations is complicated and expensive. In addition to being cautious about smooth and accurate holes when cutting metal, you must also avoid metal sticking to the drill bit. Dust and sediment are therefore removed from metal surfaces with water. Using vibration sensors in a very wet environment is problematic. For all of the reasons outlined above, we chose sound over vibration to detect broken drills in the drilling machine.


The rest of the paper is organized as follows. Section II introduces Valmet AB's dataset. Section III presents the method for data augmentation using VAE and our proposed machine failure detection system. Section IV presents the experiment result of training VAE to synthesize new drilling sounds, the classification results on the augmented dataset, and the comparison study. Section V is the conclusion.

\section{Dataset}
Four microphones with AudioBox iTwo Studio were used to capture the sounds with a sampling frequency of 96 kHz from Valmet AB's drill machines in Sundsvall, Sweden. The drilling machines at Valmet AB consist of two types of drill bits, one contains 90 drill bits, the other 120 drill bits. The sample duration for sounds in the Valmet AB dataset is very short, around 20.83~ms and 41.67~ms corresponding to 2000 sample points. The original Valmet dataset includes two categories: the "Anomaly" class (67 anomalous sounds) and the "Normal" class (67 normal sounds). "Anomaly" class includes all sounds recorded when the drill bit was broken. The "Normal" class includes sounds recorded when the drilling machine was operating properly. 

\section{Methodology}
\subsection{Variational Autoencoder}
A variational autoencoder (VAE)~\cite{kingma2014autoencoding},~\cite{rezende2014stochastic} is the architecture that belongs to the field of probabilistic graphical and variational Bayesian. A VAE is composed of an encoder and a decoder (Figure \ref{fig:Figure1}). The purpose of a VAE model isn't to replicate input sounds but to generate random variations on this input from a continuous space. The most important part of VAE is the continuous latent space that makes interpolation easier. VAE first takes an input sound and creates a two-dimensional vector (mean and variance) from a random variable (the encoding). A sampled encoding is obtained from this vector and passed to the decoder.  The decoder decodes this 1D vector and recreates the original sound. The decoder is capable of learning from all nearby points on the same latent space because encodings are generated from distributions with the same mean and variance as inputs.

\begin{figure}[!ht]
\centering
  \includegraphics[width=0.95\linewidth]{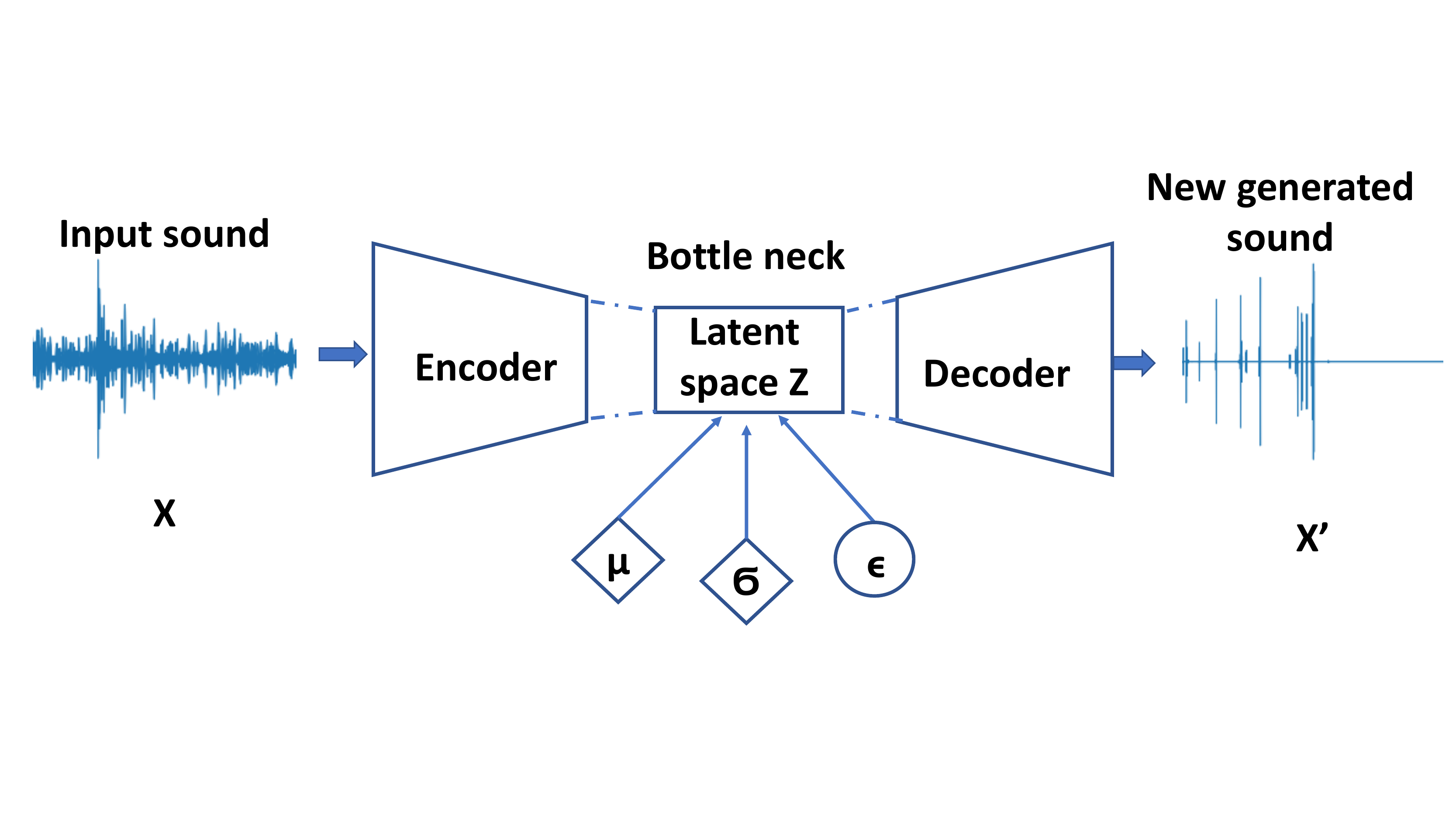}
  \caption{Autoencoder.}
  \label{fig:Figure1}
\end{figure}

The divergence between probability distributions was controlled by Kullback–Leibler divergence. Kullback–Leibler divergence is minimized by optimizing mean and variance to closely resemble that of the target mean and variance. 

\subsection{Using VAE for drilling sound synthesis}
We created a VAE to generate new drilling sounds from the sounds in the original dataset. VAE differs from traditional autoencoders in that it does not reconstruct the input using the encoding-decoding process. VAE instead imposes a probability distribution on the latent space and learns it so that the decoder produces outputs. The distribution of these outputs matches the observed data. New data are generated by sampling from this distribution. In this paper, we train a VAE on sounds in our dataset and generate new sounds that closely resemble the original sounds. Figure \ref{fig:EncoderDecoder} shows the architecture of our VAE. 


\subsubsection{Encoder}
The encoder includes four one-dimensional convolutional (Conv1D) layers interspersed with four ResnetEncoder blocks, as shown in the left side of Figure \ref{fig:EncoderDecoder}. The inputs of the encoder are 1999-length vectors corresponding to 1999 sample points of each sound in the dataset. The architecture of ResNetEncoder block is shown on the right side of Figure \ref{fig:EncoderDecoder}. ResNetEncoder block includes two one-dimensional convolutional layers and two instance normalization layers. A channel's features are normalized by instance normalization. When the learning rate is adjusted linearly with batch size, it has a more stable accuracy than the batch normalization in a wide range of small-batch sizes.

\subsubsection{Decoder}
The decoder includes four ResnetDecoder blocks interspersed with four one-dimensional convolutional transpose (Conv1DTranpose) layers, as shown in Figure \ref{fig:EncoderDecoder}. The input\_shape of the decoder is the latent dimension vector from the encoder. ResNetDecoder block includes two one-demensional convolutional layers and two batch normalization layers. A transposed convolution layer has the opposite transformation as a normal convolution layer. This is achieved by transforming the latent space that has the shape of the output of a convolution into something that has the shape of its input while maintaining a connectivity pattern that is compatible with the convolution. 



\subsubsection{Reparameterization trick}
The output of the encoder is the latent distribution. The next step is the sampling step which samples the variance and the mean vectors to pass to the decoder. However, this sampling step creates a bottleneck because the back propagation can not run through a random node, the parameterization trick is usually used to address it. As a result, we approximate \(Z\) using the decoder parameters and another parameter as follows:
\begin{equation}
    Z = \mu + \sigma*\epsilon,
\end{equation}
where $\mu$ represent the mean and $\sigma$ represent the standard deviation of a Gaussian distribution. $\epsilon$ is an auxiliary variable derived from a standard normal distribution that preserves the stochasticity of \(Z\). The parameterization trick is despicted in Figure~\ref{fig:Figure1}.

\subsubsection{Loss function}
The loss function in VAE is the sum of the reconstruction loss and Kullback–Leibler loss (KL loss) \cite{MAL-056}. The reconstruction loss measures the difference between our original input sound and the decoder output sound by using the binary cross-entropy loss. 

The KL\_loss ($KL\_loss$) is calculated by optimizing the single sample Monte Carlo estimation as below: 

\begin{equation}
KL\_loss = \log p(x|Z)+\log p(Z)-\log q(Z|x)
\end{equation}

\begin{figure*}[!ht]
\centering
  \includegraphics[width=0.78\linewidth]{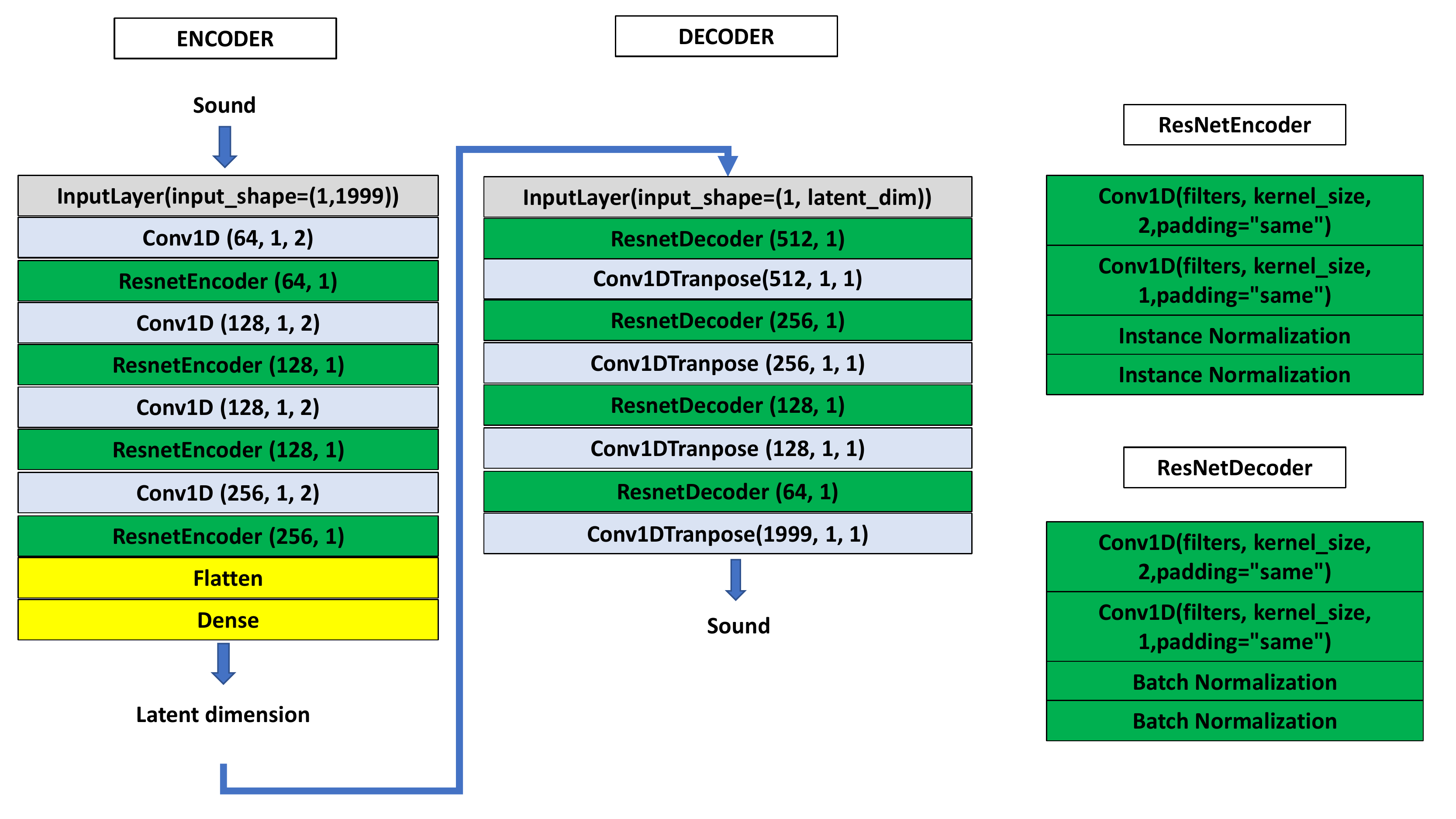}
  \caption{The variational autoencoder.}
  \label{fig:EncoderDecoder}
\end{figure*}
where $x$ represents the input sound, $Z$ represents the latent variable, $p(x)$ is the probability distribution of the sound, $p(Z)$ is the probability distribution of the latent variable, and $p(x|Z)$ is the distribution of generating data given latent variable, The probability distribution $p(Z|x)$ describes how our data is projected into latent space. $q(Z|x)$ is the inference model with \(q(Z|x) \approx p(Z|x)\)~\cite{MAL-056}.  

\subsection{The proposed machine failure detection system}
In this study, we attempted to categorize the drill sounds for the purpose of detecting drill bit failure in Valmet AB, a manufacturing in Sundsvall, Sweden. A fairly small balanced dataset for the two classes was a challenge for us when applying deep learning model. Therefore, we proposed employing VAE (a deep learning model) to generate more synthetic drilling sounds from the modest original drilling sounds as a data augmentation strategy. The original sounds were combined with the generated sounds from VAE to form the enhanced dataset. Then, a low-pass filter with a passband of 22\kern 0.16667em000 Hz and a high-pass filter with a passband of 1000 Hz were applied to this dataset as part of the preprocessing procedure. After preprocessing, these sounds were converted into Mel spectrogram and classified using the pre-trained network AlexNet. Our proposed procedure is shown in Figure~\ref{fig:systemArch}.

\begin{figure*}[!ht]
\centering
  \includegraphics[width=0.78\linewidth]{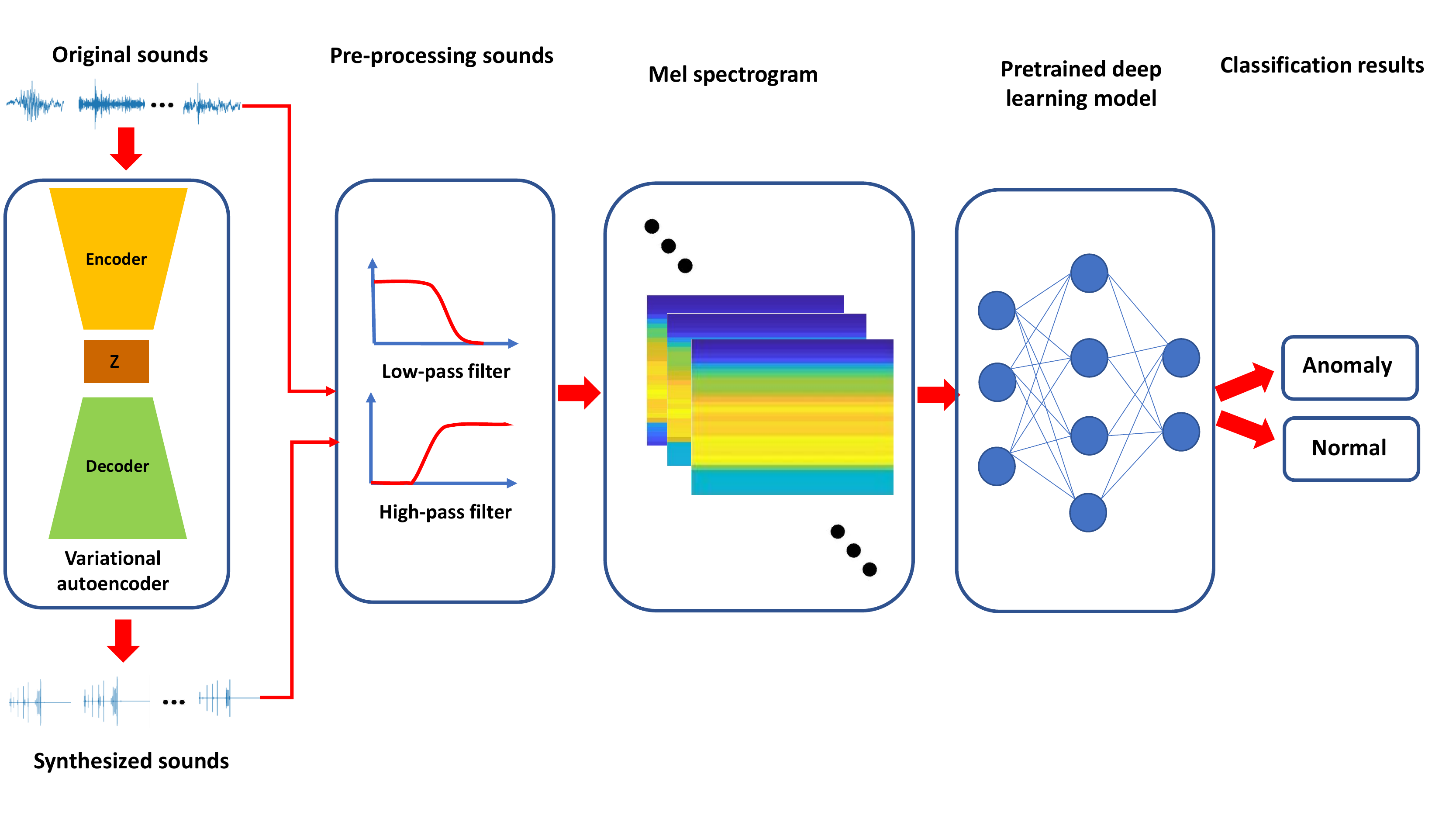}
  \caption{The proposed system architecture.}
  \label{fig:systemArch}
\end{figure*}

\section{Experimental results}
The experiment for this paper was conducted on both python and Matlab. 
\subsection{Training VAE for drilling sound synthesis}

\subsubsection{Data preparation and training VAE}
Training VAE for sound synthesis is implemented using tensorflow, numpy, pandas and librosa. The number of epoch is 20 and batch size is 8. The dimensionality of the latent space is set to 2 so the latent space is visualized as a plane. The sample rate is 96\kern0.16667em000. The training lists included 34 anomalous sounds and 34 normal drill sounds. The testing lists included 33 anomalous drill sounds and 33 normal drill sounds. The sounds are shuffled and batched using \textit{tf.data}.


The training process was iterated over the training lists. For each iteration, the sound was passed through the encoder to obtain a set of mean and log-variance for the $q(Z|x)$. This approximate posterior is then sampled using the reparameterization trick. The parameterized samples were then passed through the decoder to get the distribution $p(x|Z)$.

After training VAE on the training sounds, we generated 48 drilling sounds for the "Anomaly" class and 48 normal drilling sounds for the "Normal" class. 


\subsubsection{Measuring sound similarities between the original sound and the synthesized sound}
We hypothesis that VAE maybe reduce the noise in the synthesized data. However, by listening to the synthesized drilling sounds, the newly generated drilling sounds closely follow the pitch of the original sounds. Hence, we assessed the similarity between a sound in a testing set and a synthesized drilling sound from VAE using the cross-correlation and the spectral coherence between two sounds using Matlab.

\textbf{Cross-correlation between two sounds:}

We measured the similarities between an original sound in the anomaly class and the synthesized sound from VAE. These two sounds have the same sampling rate of 96\kern 0.16667em000 Hz but have different lengths. The original sound has a length of 41.67 ms whereas the synthesized sound has a length of around 20.83 ms. The calculation of the difference between two sounds is difficult because of the different lengths so we had to extract the common part of the two sound signals using the cross-correlation. Figure~\ref{fig:cross-correlation} shows the cross-correlation between the original sound and the synthesized sound. The value of cross-correlation coefficient is from -1.0 to 1.0. If the cross-correlation value is closer to 1, the two sounds are more similar. The high peak in the third subplot shows that the original sound correlated to the synthesized sound.
\begin{figure}[!ht]
\centering
  \includegraphics[width=1\linewidth]{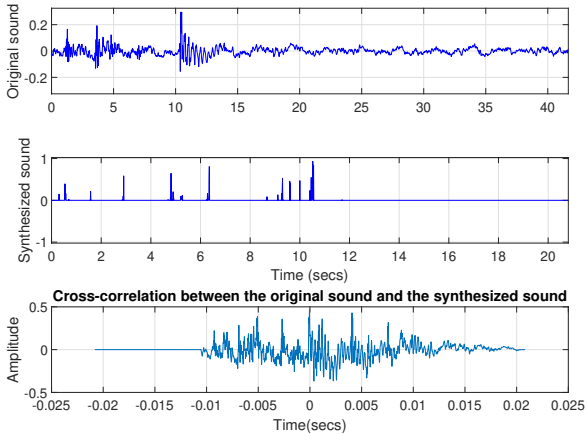}
  \caption{The cross-correlation between the original sound and the synthesized sound.}
  \label{fig:cross-correlation}
\end{figure}

\textbf{The spectral coherence between two sounds:}

Power spectrum shows how much power is present per frequency. Correlation between signals in the frequency domain is referred to as spectral coherence. The power spectrum of Figure~\ref{fig:spectral_coherence1} shows that the original sound and synthesized sound have correlated components in the range of 0 Hz and 6890 Hz. 
\begin{figure}[!ht]
\centering
  \includegraphics[width=1\linewidth]{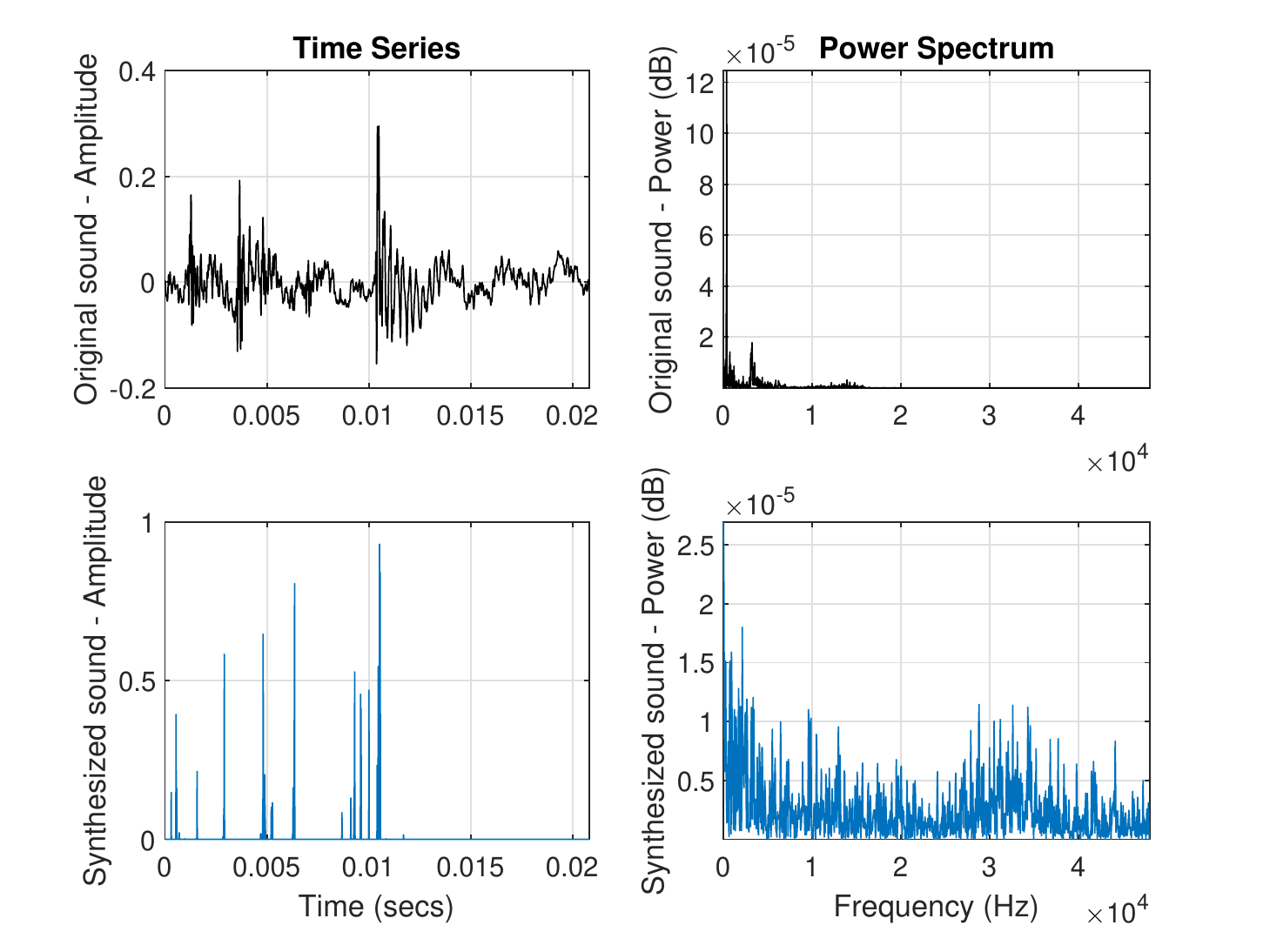}
  \caption{The power spectrum.}
  \label{fig:spectral_coherence1}
\end{figure}

We show the coherence estimate in the first subplot of Figure~\ref{fig:spectral_coherence2}. Coherence values range from 0 to 1. For example, we marked three highest peak of the coherence values as 0.494153, 0.55872, and 0.377155, as shown in Figure~\ref{fig:spectral_coherence2}. The higher the coherence, the more correlated the corresponding frequency components are. There are approximately 45 degrees of phase lag between the 1125 Hz components and the 4312.5 components. The phase lag between the 9187.5 Hz components is approximately 38 degrees.

\begin{figure}[!ht]
\centering
  \includegraphics[width=1\linewidth]{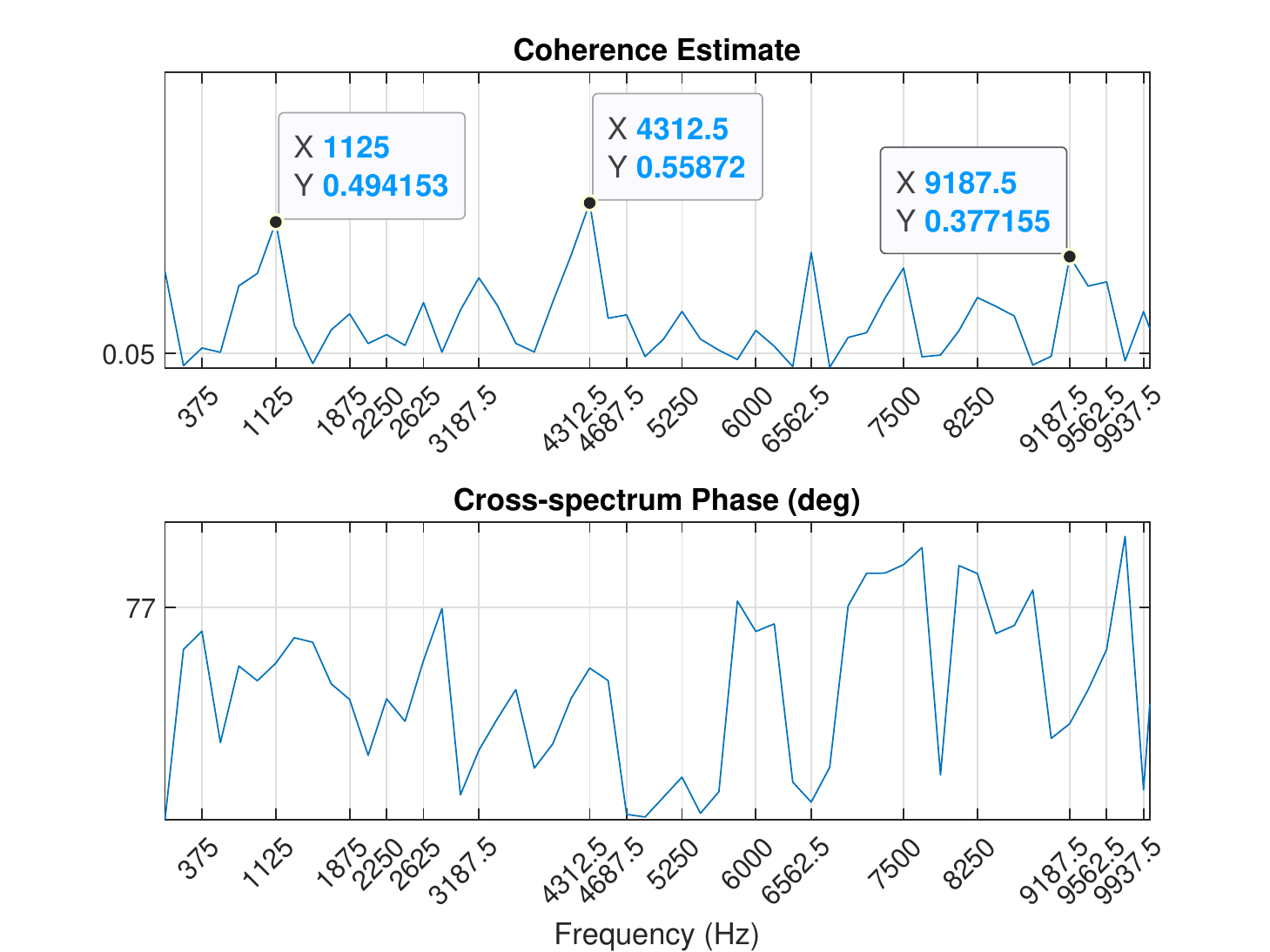}
  \caption{The spectral coherence.}
  \label{fig:spectral_coherence2}
\end{figure}

\subsection{Classification result using original and synthesized sounds}
We mixed the orginal dataset that contained 67 sounds for each class with the generated dataset that contained 48 synthesized sounds for each class. We obtained the augmented dataset that included two classes, "Normal" class (115 sounds) and "Anomaly" class (115 sounds).
\subsubsection{Sound pre-processing}
The sample rate of sounds in our augmented dataset is 96\kern 0.16667em000, but the sound of a drill typically ranges from roughly 1000 to 22\kern 0.16667em000 Hz. The low-pass filter are used to eliminate frequencies higher than 22\kern 0.16667em000 Hz. The frequency below 1000 Hz was then removed using a high pass filter.

\subsubsection{Convert sounds to Mel spectrograms}
We converted sounds into Mel spectrograms using the Mel spectrums of 1999-point periodic Hann windows with 512-point overlap. 1999-point FFT was used to convert to the frequency domain and pass this through 32 half-overlapped triangular bandpass filters. 
Figure~\ref{fig:MelSpec}a shows the Mel spectrogram of a sound without applying low-pass filter and high pass filter. Figure~\ref{fig:MelSpec}b shows the Mel spectrogram after applying low-pass filter and high pass filter. 


\begin{figure}[hpt!]
    \centering
    \subfigure[]{\includegraphics[width=0.49\linewidth]{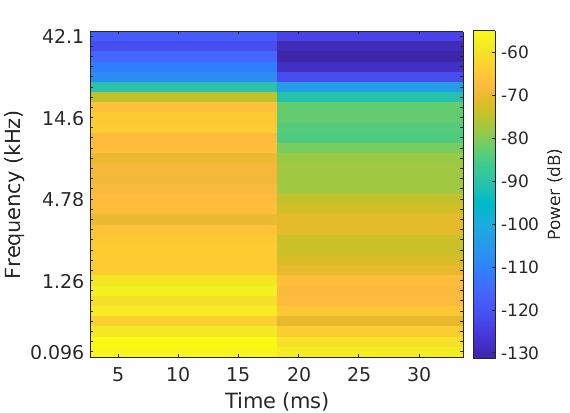}} 
    \subfigure[]{\includegraphics[width=0.49\linewidth]{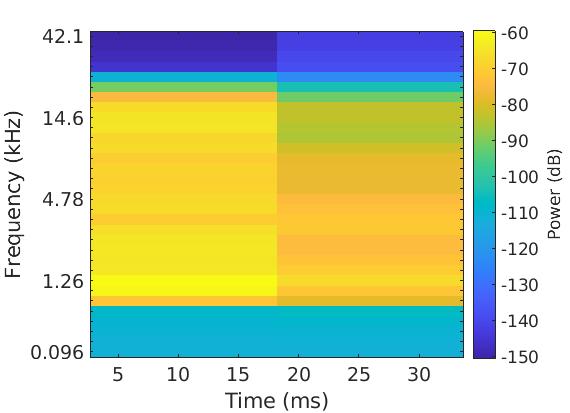}} 
    \caption{The Mel spectrogram of an original sound in the "Anomaly" class: (a) without applying low-pass filter and high-pass filter  (b) apply low-pass filter and high-pass filter.}
    \label{fig:MelSpec}
\end{figure}


\subsubsection{Classification result}
To demonstrate the effectiveness of using VAE for generating new sounds, we proceed to classify two classes in the augmented dataset (mix between original sounds and synthesized sounds) using several pre-trained networks as in Table~\ref{tablecompaAccuracy}. This experiment is conducted on Matlab 2021a. In overall, pre-trained AlexNet shows the best accuracy (94.12\%) on our dataset. AlexNet is a simple CNN with only eight layers that was trained to classify 1000 objects from the ImageNet dataset. 

\begin{table}[h]
\centering
\caption{The accuracy of different pre-trained models on the augmented dataset.}
\label{tablecompaAccuracy}
\begin{tabular}{l|c}
\hline
\textbf{Pre-trained model}                  & \textbf{Accuracy (\%)} \\ 
\hline
GoogLeNet               & 89.71                       \\ \hline
\textbf{AlexNet}         & \textbf{94.12}                       \\ \hline
VGG16   & 88.24                     \\ \hline
VGG19         & 85.29                       \\ \hline
Efficientnetb0         & 86.76                       \\ \hline
\end{tabular}
\end{table}

Because after training VAE to generate new drilling sounds, our augmented dataset is extended but still small. Hence, we fine-tuned a common AlexNet with transfer learning instead of training a new CNN from scratch. The learned features from pre-trained AlexNet can easily and quickly transfer to a new task of classification the "Normal" class and "Anomaly" class in the augmented dataset. Sounds in the augmented dataset were converted into Mel spectrogram, as described above.  

We utilized 70\% of Mel spectrograms (162 sounds) for training and 30\% for validation (68 sounds). The input size of AlexNet is 227-by-227-by-3, where 3 represents three R, G, B channels.  The fully connected layer and classification layer of AlexNet were used to classify 1000 classes. To fine-tuned AlexNet to classify only two classes in our augmented dataset, the fully connected layer was replaced with a new fully connected layer that has only two outputs. the classification layer was replaced with a new classification layer that has two output classes. 

We augmented the training set (162 sounds) by flipping the Mel spectrograms along the x-axis, translating them via the x-axis and the y-axis randomly in the range of (-30, 30) pixels, and scaling them via the x-axis and the y-axis randomly in the range of (0.9, 1.1).  

For the training option, mini-batch size was 10, the validate frequency (\(valFreq\)) is calculated as below:
\begin{equation}
valFreq = training\_file/ mini\_batch\_size
\end{equation}
where training\_file is the number of the training file. Max epochs were 160, the initial leaning rate is 3e-4, shuffle every epoch.

The overall accuracy when fine-tuning AlexNet on our augmented dataset reaches 94.12\%. The confusion matrix is shown in Figure \ref{fig:ConfusionMatrix_ImgAugmentation} whereas the accuracy when detect "Anomaly" class reaches 97.06\%.

\begin{figure*}[hpt!]
    \centering
    \subfigure[]{\includegraphics[width=0.35\linewidth]{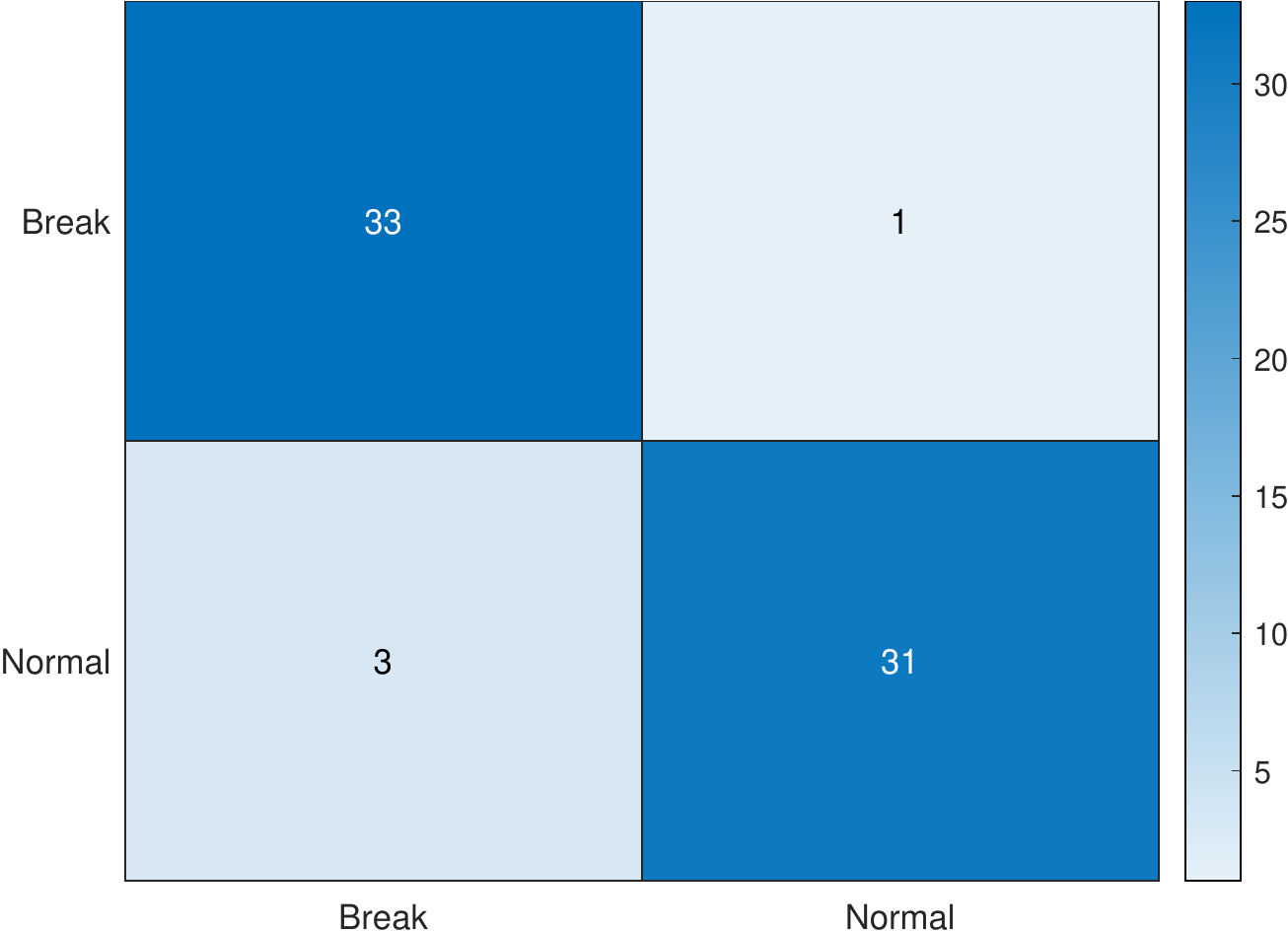}} 
    \subfigure[]{\includegraphics[width=0.35\linewidth]{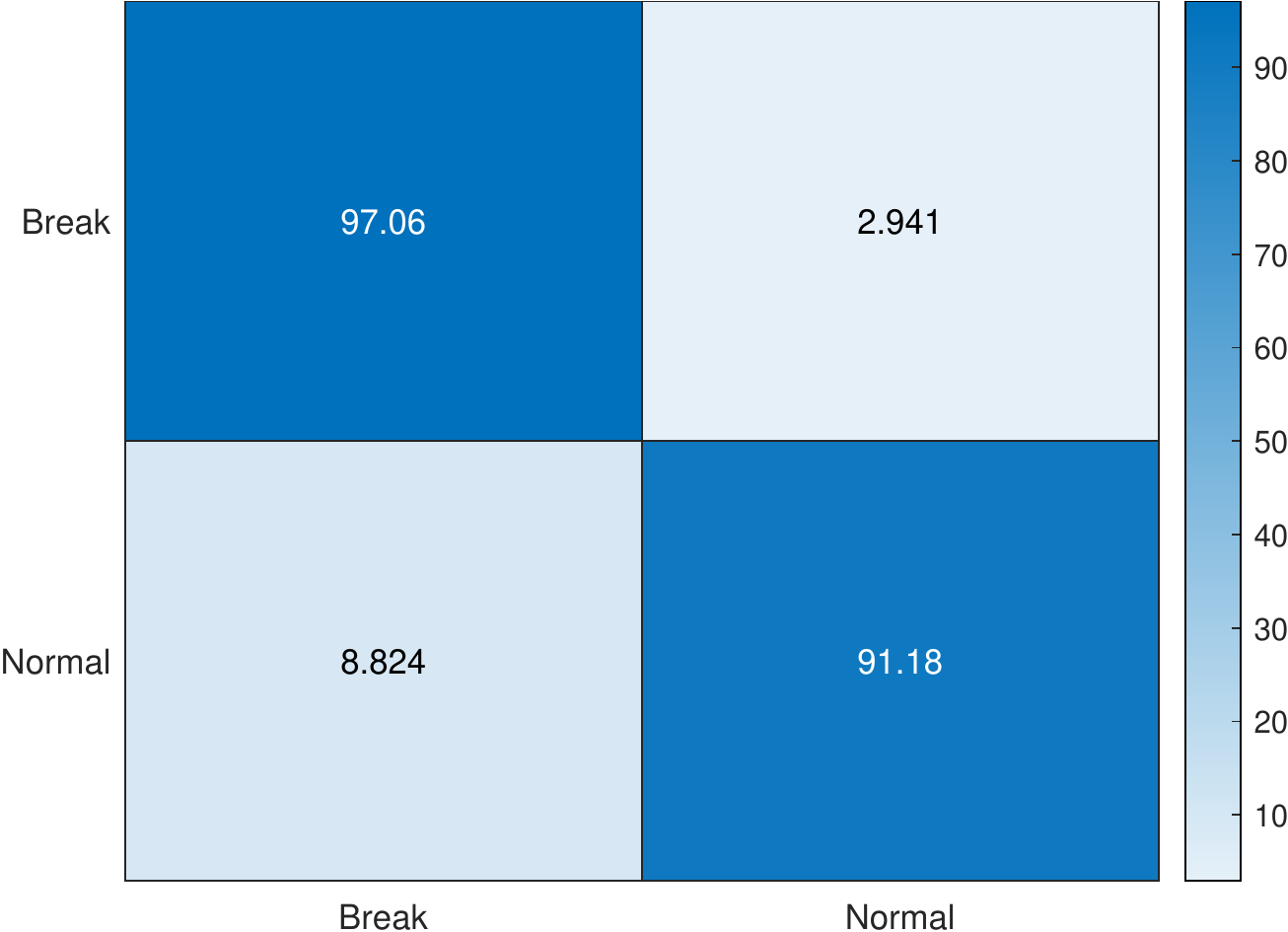}} 
    \caption{The confusion matrix when fine-tuning AlexNet to classify "Normal" and "Anomaly" classes on the augmented dataset: (a) in number (b) in percentage.}
    \label{fig:ConfusionMatrix_ImgAugmentation}
\end{figure*}

\subsection{Analyzing the original dataset with pre-trained Alexnet}
This section compares the accuracy of the pre-trained network AlexNet on the original dataset and on the augmented dataset. We trained pre-trained AlexNet with the same training option on the original drilling dataset, the accuracy reached 87.5\%, as shown in Table ~\ref{tablecompare1}. Meanwhile, we found that when we combined the original drilling sounds and synthesized sounds derived from VAE (the augmented dataset), the classification accuracy was 94.12\%, which is 10.15\% higher than when we used the small original dataset. The confusion matrix for the original dataset is shown in Figure~\ref{fig:ConfusionMatrix_original}.


\begin{table}[h]
\centering
\caption{The comparison of AlexNet on the original dataset and the augmented dataset. No. of sounds indicates the number of sounds.}
\label{tablecompare1}
\begin{tabular}{l|l|c}
\hline
\textbf{Dataset}              & \textbf{No. of sounds} & \textbf{Accuracy (\%)} \\ \hline
The original drilling dataset & 134                   & 87.5                      \\ \hline
The augmented dataset         & 230                   & 94.12                  \\ \hline
\end{tabular}
\end{table}


\begin{figure*}[hpt!]
    \centering
    \subfigure[]{\includegraphics[width=0.35\textwidth]{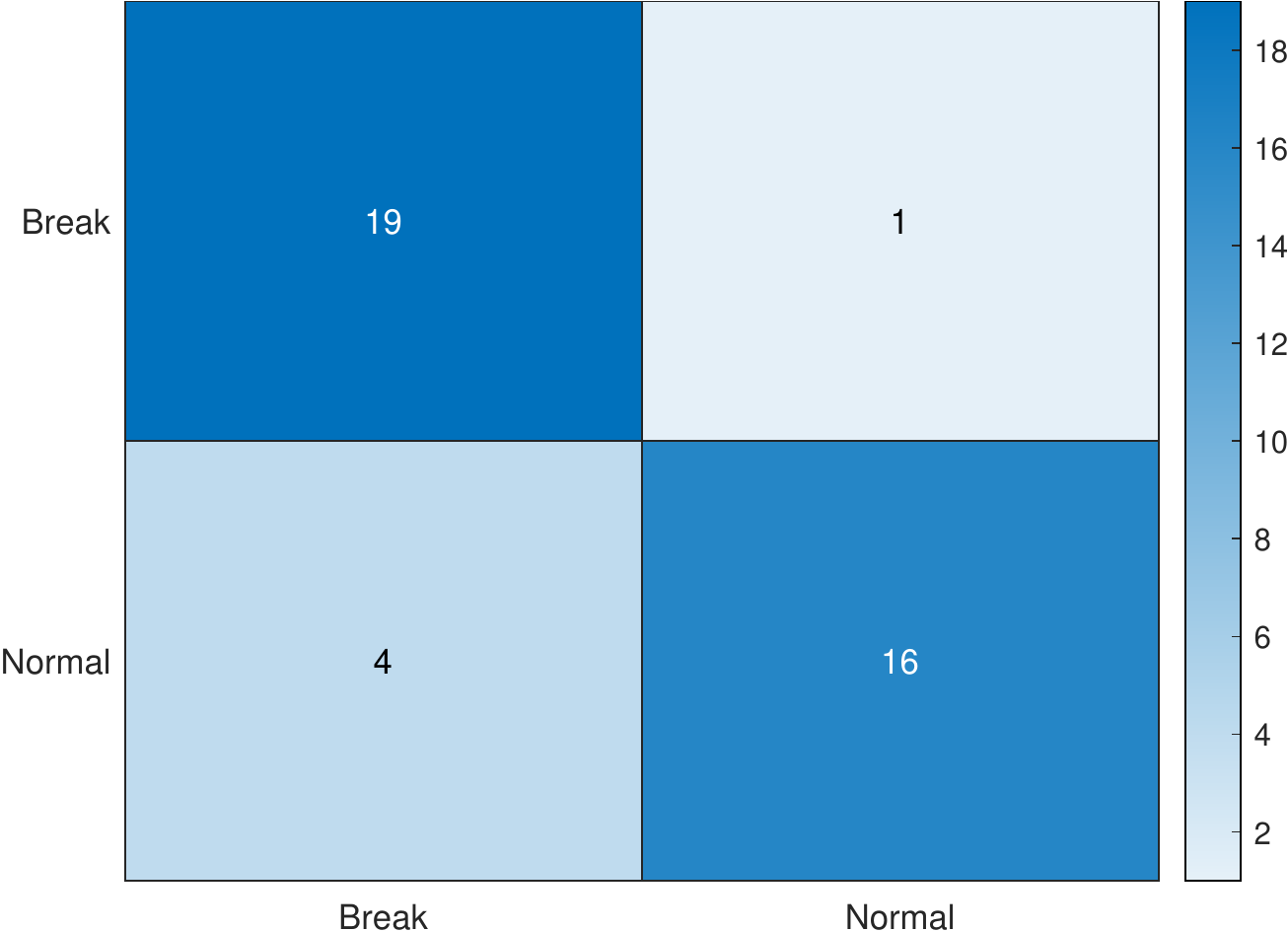}} 
    \subfigure[]{\includegraphics[width=0.35\textwidth]{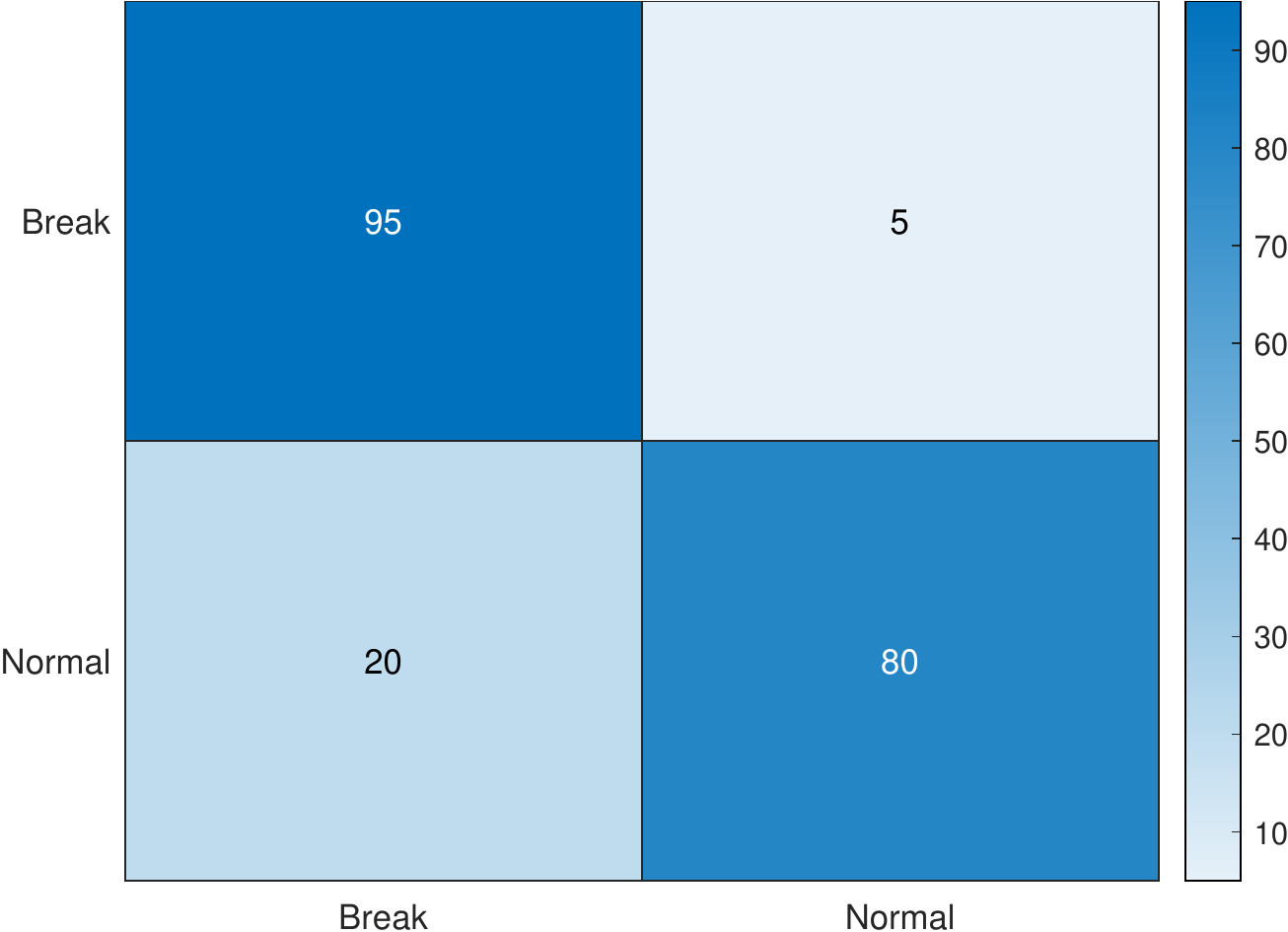}} 
    \caption{The confusion matrix when fine-tuning AlexNet to classify "Normal" and "Anomaly" classes on the original dataset.: (a) The confusion matrix in number  (b) The confusion matrix in percentage (\%).}
    \label{fig:ConfusionMatrix_original}
\end{figure*}

\section{Conclusion}
Drill failure detection is critical for the industry to reduce drilling machine downtime. The unavailability of drilling sounds in the dataset, on the other hand, made it impossible to adequately train the failure detection model. As a result, employing VAE to synthesize new drilling sounds from existing drilling sounds can help to supplement the small sound dataset. A case study is presented here that uses 134 drill sounds classified into two categories (normal and anomalous) in a Valmet AB dataset. We used a VAE to synthesize 96 new drilling sounds (48 normal sounds and 48 anomaly sounds) from the original drilling sounds. VAE may, although, minimize the noise in synthetic data. However, listening to the synthetic drilling sounds reveals that the newly generated drilling sounds roughly match the pitch of the original sounds. The newly synthesized sounds and original sounds were combined to create the augmented sound dataset to train a discrimination model. We fine-tuned AlexNet to classify Mel spectrograms of sounds in the augmented dataset. The result is also compared to fine-tuned AlexNet on the original dataset. The overall classification accuracy reached 94.12\%. The result is promising to build a real-time machine fault detection system in the industry.

\section*{Acknowledgment}
This research was supported by the EU Regional Fund, the MiLo Project (No. 20201888), and the Acoustic sensor set for AI monitoring systems (AISound ) project. The authors would like to thank Valmet AB for providing the drill sound dataset. 

The computations/data handling was enabled by resources provided by the Swedish National Infrastructure for Computing (SNIC) at the SNIC center is partially funded by the Swedish Research Council through grant agreement no. 2018-05973.

\Urlmuskip=0mu plus 1mu\relax
\bibliographystyle{IEEEtran}
\bibliography{MyCollection.bib}
\end{document}